\documentstyle{BSAXwork}

\input epsf.sty

\def \AAP #1 #2 {{\em Astron. Astrophys.\/} {\bf #1}, #2}
\def \AJ #1 #2 {{\em Astron. J.\/} {\bf #1}, #2}
\def \APJ #1 #2 {{\em Astrophys. J.\/} {\bf #1}, #2}
\def \APJL #1 #2 {{\em Astrophys. J. Lett.\/} {\bf #1}, L#2}
\def \MN #1 #2 {{\em Mon. Not. R. Astr. Soc.\/} {\bf #1}, #2}
\def \PASA #1 #2 {{\em Publ. Astron. Soc. Australia\/} {\bf #1}, #2}
\def \PASJ #1 #2 {{\em Publ. Astron. Soc. Japan\/} {\bf #1}, #2}

\begin{opening}

\title{ On particle acceleration in turbulent shear boundary layers of
 relativistic jets }
\author{M. Ostrowski$^{1}$, \L. Stawarz}
\institute{Obserwatorium Astronomiczne, Uniwersytet Jagielloñski, ul. Orla 171,
 30-244 Kraków, Poland\\
$^1$mio@oa.uj.edu.pl }
\date{}
\end{opening}

\begin{document}

\oddpagefooter{}{}{}
\evenpagefooter{}{}{}
\medskip

\begin{abstract}
We discuss processes accelerating cosmic ray electrons at boundaries of
 relativistic jets. The resulting spectrum is expected to posses two
 characteristic components: a power-law distribution at lower energies and a
 harder high energy component preceding the high energy cut-off. An example of high energy
 spectrum from such electron distribution is presented, including the
 synchrotron and the inverse-Compton components. A characteristic feature of the
 synchrotron spectrum of such electrons is its highest frequency part which occurs
 above the power-law fitted to the low frequency spectral range.
\end{abstract}

\medskip

\section{Introduction}
Radio and optical polarimetry of the large scale jets in radio galaxies usually
 suggest strong velocity shearing effects at the jet edges, resulting in a
 magnetic field being parallel to the jet axis (see, e.g., Perlman et al. 1999
 for M 87, Attridge et al. 1999 for 1055+018). Acceleration of radiating
 particles within a boundary region is required to explain the observed jet
 radial profile (e.g. in M87, Owen et al. 1989), or details of the spectral
 maps, like flattening of the radio spectrum near the jet surface in Mkn 501
 (Edwards et al. 2000). One should mention, that the HST studies of the optical
 counterparts of large scale radio jets emphasize the necessity of the
 continuous electron reacceleration in a jet body (and not only in shock
 regions), as the lifetime of the synchrotron electrons is often much shorter
 than the light-travel time along the optical structure (cf. Jester et al. 2001 for
 the case of 3C 273). On the other hand, the broad band spectral properties of
 BL Lacs and FR I radio galaxies indicate a jet stratification. By comparing the
 multiwavelength observations of these two classes of AGNs in a framework of a
 unification scheme, Chiaberge et al. (2000) found evidences for a significant
 boundary layer emission, dominating the radiative jet output in FR Is. The
 inferred velocities of such boundary regions are significantly lower as
 compared to velocities of central spines, but are still relativistic, in order
 to explain anisotropic emission observed in the FR Is' cores.

3D hydrodynamical simulations of relativistic jets reveal a highly turbulent
 cocoon and a shear layer surrounding the jet spine (Aloy et al. 1999). The
 reason for the presence of a turbulent medium effects in these simulations is
 numerical viscosity, which is not a real physical process. However, theoretical
 considerations support the model of a turbulent medium at the jet edges
 (cf. Bicknell \& Melrose 1982). Such regions are therefore the promising places for
 particle acceleration, as discussed previously by Ostrowski (1998, 2000). The
 jet boundary layer acceleration, when applied to protons, was considered to
 provide ultra high energy cosmic rays, influence the jet dynamics, and
 significantly increase a pressure in the radio lobes of FR II sources
 (Ostrowski \& Sikora 2001). Electrons accelerated at the jet boundary were
 considered to provide important contribution for the jet radiative output,
 including the large scale X-ray emission and its possible time (spatial)
 variation (Stawarz \& Ostrowski 2002, 2001).

\section{Radiating boundary layer}
For illustration let us consider a relativistic large scale jet consisted of a spine surrounded
 by the boundary layer with a velocity shear. In order to specify the velocity
 structure, we assume a uniform flow Lorentz factor $\Gamma_j$ inside the jet
 spine, and a linear radial profile $1 < \Gamma < \Gamma_j$ within the boundary
 transition region. Below we put the thickness of the boundary layer to be of
 order of the jet radius, $D \sim 1 \, {\rm kpc}$. Furthermore, we assume, that
 due to shearing effects the magnetic field is parallel {\it on average} to
 the jet axis inside the highly turbulent boundary layer, and that its intensity
 is the same as the one estimated for the spine, $B = 10^{-5} \, {\rm G}$.

The electrons injected into the jet boundary region can be accelerated due to
 {\it cosmic ray viscosity} connected with the flow velocity radial gradient
 within the shear layer and due to stochastic Fermi acceleration in the
 turbulent medium. For the assumed large scale jet parameters (with $D \gg r_g$,
 where $r_g$ is the electron gyroradius) the later mechanism acts more
 efficiently and, hence, the electrons gain energy mostly due to scattering on
 the long wavelength magnetic field irregularities moving with the velocity
 comparable to the Alfv\'en velocity $V_A$. The acceleration time scale for this
 process is roughly $T_{acc} \sim \zeta \, r_g \, c \, V_A^{-2}$, where $\zeta$
 is the ratio of the energy density of the regular magnetic field to the
 turbulent one. For the jet dynamicaly dominated by nonrelativistic protons,
 $V_A$ is a few orders of magnitudes smaller than the flow velocity $U \sim c$.
 Below we assume $V_A \sim 0.01 \, c$, what corresponds to the cold proton
 number density $n_p \sim 10^{-4} \, {\rm cm^{-3}}$ (cf. Sikora \& Madejski
 2000).

Radiative losses in the large scale jets are due to the synchrotron emission and
 comptonization of cosmic microwave background (CMB) photons and synchrotron
 core emission (AGN) illuminating the large scale jet from behind (cf. Celotti
 et al. 2001). In the Thomson regime one can write $T_{loss} \sim 3 \, m_e \, c
 / 4 \, \sigma_T \, \gamma \, u$, where $u = u_B + u_{cmb} + u_{agn}$ is the
 energy density of the magnetic field and the seed photons in the source frame.
 The latter two represent anisotropic radiation fields which depend on the flow
 Lorentz factor of the emitting region, and hence -- in our case of a shear
 layer -- on the distance from the jet axis. For simplicity, we limit our
 calculation to the upper limits of both $u_{cmb} $ and $u_{agn}$, and neglect
 cosmological redshift corrections. Then the energy density of the CMB
 background is $u_{cmb} = a \, T_{cmb}^4 \, \Gamma^2 < a \, T_{cmb}^4 \,
 \Gamma_j^2$, and the energy density of the core emission with intrinsic
 luminosity $L'_{agn}$ is $u_{agn} = L'_{agn} \, \Gamma_0^2 \, / 4 \pi c \, z^2
 \, \Gamma^2 < L'_{agn} \, \Gamma_0^2 \, / 4 \pi c \, z^2$, where $\Gamma_0$ is
 the bulk Lorentz factor of the (sub-) parsec scale jet, and $z$ is the distance
 from the galactic nucleus. The electrons' escape from the shear layer with the assumed
 magnetic field configuration proceeds through the cross-field diffusion process. The
 appropriate time scale for this process can be estimated as $T_{esc} \sim D^2 \,
 \kappa_{\bot}^{-1}$, where $\kappa_{\bot} \sim \eta \, r_g \, c / 3$ is an
 effective cross-field diffusion coefficient, with the numerical scaling factor
 $\eta \leq 1$. $T_{esc}$ is much longer than the time scale for radiative
 losses, $T_{loss}$. As a result, the electrons pile-up at highest energies below the
 maximum energy $\gamma_{eq} \, m c^2$, where $T_{acc} \approx T_{loss}$. Below we will
 model the formed hard component in the electron spectrum with a monoenergetic bump at
 the highest energy ($\propto \delta ( \gamma - \gamma_{eq})$, cf. Ostrowski 2000).
 For the parameters $L'_{agn} \sim 10^{43} \, {\rm erg/s}$,
 $\zeta = 1$ and $\Gamma_0 \sim \Gamma_j \sim 10$, the maximum energy the
 electrons can reach due to turbulent acceleration in a Bohm limit -- at distances
 $z \geq 10 \, {\rm kpc}$ where $u_{agn} < u_{cmb}$ -- corresponds to the
 electron Lorentz factor $\gamma_{eq} \sim 10^8$.

\begin{figure}
\epsfysize=9cm
\hspace{0.0cm} \vspace{0.0cm}\epsfbox{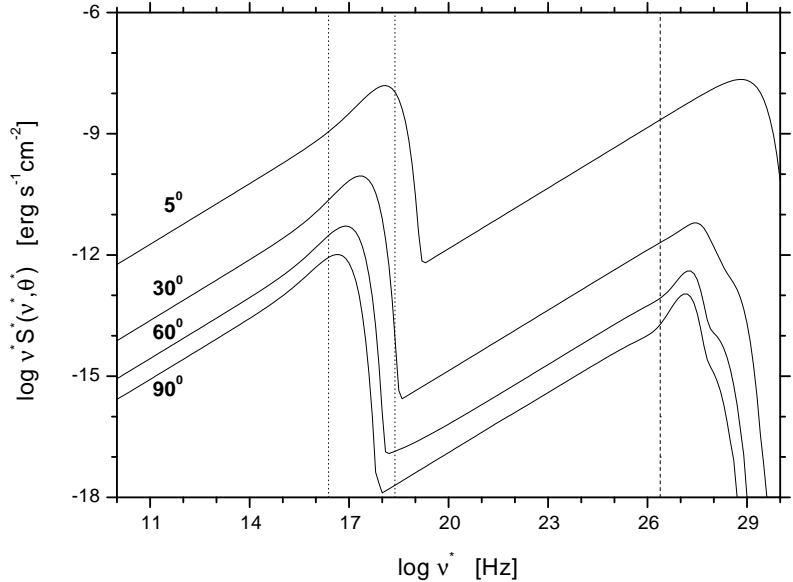}
\caption[h]{ The observed spectral energy distributions of the radiation
 generated within the shear boundary layer for $\gamma_{eq} = 10^8$ and
 different jet inclinations, as indicated near the respective curves. The
 presented spectra are integrated over the assumed linear flow Lorentz factor
 profile with $D \sim 1 \, {\rm kpc}$ and $\Gamma_j \sim 10$. A distance to the
 observer is assumed to be 10 Mpc, and the absorption of VHE $\gamma$-rays
 during the propagation to the observer is neglected. The dotted vertical lines
 indicate the Chandra energy band, 0.1 keV - 10 keV. The dashed vertical line
 indicates the observed photon energy 1 TeV.  }
\end{figure}

In the case of efficient particle injection, the acceleration process acting
 continuously within the whole jet boundary layer leads to the distribution
 approximated here as a flat power-law spectrum $n_e (\gamma) \propto
 \gamma^{- \sigma}$ finished with the pile-up bump at the maximum energies
 near $\gamma_{eq}$. The detailed evolution of the highest energy electrons needs
 careful analysis of the momentum diffusion of the accelerating particles within
 turbulent layer with velocity and density gradients (work in preparation).
 Below, in order to discuss main consequences of the boundary layer acceleration
 for the radiative jet output, we consider a case of forming the stationary two-component
 electron energy distribution with $\sigma \sim 2$ and the high energy bump modeled
 as a monoenergetic peak (c.f. discussion in Stawarz \& Ostrowski 2002). The
 observed multiwavelength radiation of such electrons, including synchrotron
 radiation, synchrotron self-Compton emission and external Compton scattering of
 CMB photons is plotted on figure 1 for different jet inclinations. In order to
 find the normalization constants for the particle distribution, we assumed the
 energy equipartition between the magnetic field and each one of the electron spectral
 components. The presented spectra are integrated over the assumed bulk flow
 Lorentz factor profile within the boundary layer, with the appropriate beaming
 pattern for synchrotron/self-Compton and external Compton radiation (Dermer
 1995).

\section{Discussion - X-ray emission of the large scale jets}

After Chandra discovery that the large scale jets in quasars and radio galaxies
 are strong X-ray emitters, the question of the origin of such radiation became
 an important issue. For the jets observed at small angles, the inverse Compton
 scattering of the CMB photons by the low-energy tail of the non-thermal
 electron distribution flowing with large bulk Lorentz factor is the most likely
 explanation (Tavecchio et al. 2000). However, strong beaming effects exclude
 this process in case of the jets in radio galaxies. Celotti et al. (2001)
 proposed instead, that the X-ray radiation detected from such objects can be
 explained as the SSC emission of the slower (and therefore less beamed)
 boundary layer electrons, or as the inverse Compton scattering of the core
 emission. This, however, requires a departure from the energy equipartition
 between the magnetic field and radiating particles, at least if the X-ray flux
 is of the same order of magnitude as the synchrotron one. As illustrated on
 figure~1, the boundary layer acceleration at the tens-of-kpc scale jets can
 generate high energy synchrotron radiation peaking at the keV energy band.
 Luminosity of such radiation is high enough to account for the Chandra
 observations {\it without} departures from the equipartition condition (Stawarz
 \& Ostrowski 2002). Continuous acceleration acting within the whole boundary
 layer volume compensates radiative losses of high energy electrons, and
 therefore the diffusive character of their X-ray emission is a natural
 consequence of the presented model (but of course any compressive perturbation
 -- a shock -- in the jet can disturb this smooth intensity variation). One should
 note that recent HST observations of the jet in 3C 273 (Jester et al. 2002)
 with a smooth spectral transition from optical, through UV, up to X-ray frequencies
 are consistent with our model.

The low energy (radio-to-optical) component of the boundary layer synchrotron
 emission can affect the jet-counterjet brightness asymmetry measurements,
 `hiding' the highly relativistic spine. Thus the bulk Lorentz factor of the jet
 spine at large scales can be comparable to the one at VLBI scales. This has
 several consequences for the jet energetics (cf. Ghisellini \& Celotti 2001).

\acknowledgements
We are grateful to Marek Sikora for his help and discussions. The present work
 was supported by Komitet Bada\'{n} Naukowych through the grant BP
 258/P03/99/17.

\end{document}